# Optical properties of Fermi polarons in a GaInP/MoSe$_2$ monolayer heterostructure

Hangyong Shan[1], Max Waldherr[2], Diksha Diksha[1], Ghada Missaoui[1], Seyma Esra Atalay[1], Martin Zinner[2], Ana Maria Valencia[1], Kenji Watanabe[4], Takashi Taniguchi[5], Seth Ariel Tongay[3], Caterina Cocchi[1,9], Brendan C. Mulkerin[6], Jesper Levinsen[6], Francesca Maria Marchetti[7,8], Meera M. Parish[6], Niklas Nilius[1], Christian Schneider[1,†], Sven Höfling[2]

[1]*Institute of Physics, Faculty V, and Center for Nanoscale Dynamics (CeNaD), Carl von Ossietzky University Oldenburg, 26129 Oldenburg, Germany*

[2]*Technische Physik and Wilhelm-Conrad-Röntgen-Research Center for Complex Material Systems, Universität Würzburg, D-97074 Würzburg, Am Hubland, Germany*

[3]*Materials Science and Engineering, School for Engineering of Matter, Transport, and Energy, Arizona State University, Tempe, Arizona 85287, USA*

[4]*Research Center for Electronic and Optical Materials, National Institute for Materials Science, 1-1 Namiki, Tsukuba 305-0044, Japan*

[5]*Research Center for Materials Nanoarchitectonics, National Institute for Materials Science, 1-1 Namiki, Tsukuba 305-0044, Japan*

[6]*School of Physics and Astronomy, Monash University, Victoria 3800, Australia*

[7]*Departamento de Física Teórica de la Materia Condensada, Universidad Autónoma de Madrid, Madrid 28049, Spain*

[8]*Condensed Matter Physics Center (IFIMAC), Universidad Autónoma de Madrid, 28049 Madrid, Spain*

[9]*Friedrich-Schiller University Jena, Institute for Condensed Matter Theory and Optics, 07743 Jena, Germany*

[†]*Corresponding author. Email: christian.schneider@uol.de*

**Abstract:**

Engineering optical properties, such as luminescence purity and charge transfer, is crucial for harnessing the application potential of atomically thin transition metal dichalcogenides (TMDCs). While electrostatic gating is widely applied to gain charge control in TMDC monolayers, charge transfer can also be engineered via coupling of TMDC monolayers at semiconductor III/V, organic, or van der Waals interfaces. This confers great advantages, such as ease in implementation and compatibility in device integration. Here, we shed light on the optical properties of many-particle complexes emerging at the GaInP/MoSe$_2$ interface as a highly relevant material combination to manipulate the optical properties of TMDCs in integrated photonic devices. Our study verifies its nature as a type II hetero-interface, which bears the feasibility to display disorder-free photoluminescence. Through optical absorption measurements, we verify that the charged complexes acquire substantial oscillator strength. Furthermore, temperature-dependent photoluminescence, supported by a microscopic theory framework, evidences the suppression of the characteristic carrier recoil effect that was previously observed in the photoluminescence of trions in TMDCs. These phenomena allow us to identify the optical signatures at the TMDC-GaInP interface as Fermi polaron quasiparticle resonances, which are of high importance in researching Bose-Fermi mixtures in condensed matter systems.

**Introduction**

Atomically thin transition metal dichalcogenides (TMDCs) have emerged as an excellent platform to study fundamental effects of light-matter coupling and exciton physics in low dimensions[1], and to explore correlated quantum states of matter[2,3]. Being van der Waals materials, TMDC monolayers and heterostructures are extremely versatile regarding their integration with electronic[4] and photonic[5] device structures, which explains their tremendous application potential. However, due to their ultimate thinness, the structural surrounding of a TMDC monolayer, such as the choice of the substrate material, widely influences its optoelectronic properties via several effects, including photonic coupling, dielectric screening, disorder, and charge transfer[6-10]. Consequently, it is of significant interest to shed light on the optical properties of device-relevant combinations of TMDCs with commonly used materials in optoelectronic research, such as III/V semiconductors.

A variety of hybrid III/V-TMDC structures has already been employed in photonic applications, including hybrid polariton lasers[11] and other nanophotonic devices[12], where excitonic effects in TMDCs dominate the optical response. Further studies have characterized excitons at the TMDCs-semiconductor interface from a fundamental perspective[9]. Besides excitons, charged complexes such as trions are observed in TMDCs due to strong Coulomb interactions[13,14]. Trion resonances can become very prominent when TMDC monolayers are integrated with carrier-donating substrates[15,16], which also attract particular interest due to their fermionic nature[17]. When the free carrier density becomes greater than the exciton density, excitons can be regarded as quantum impurities immersed in a Fermi sea. Within this picture, excitons and trions evolve into repulsive (RP) and attractive Fermi polarons (AP) respectively, where excitons either repel or attract the surrounding Fermi particles[18-22]. The optical properties of Fermi polarons in atomically thin semiconductors – such as their emission frequency, dephasing properties, and oscillator strength – strongly depend on the doping density[19,21,22], which can be modified either by electric gates or various substrates.

The Fermi polaron represents a ubiquitous scenario in physics, with realizations spanning a wide range of platforms, from ultracold atomic gases[23], to high-energy and condensed-matter systems[24]. Major advances in the context of atomically thin semiconductors include the observation of a strongly enhanced nonlinear optical response[25], as well as the controlled manipulation of exciton–polaron states using external electric[26,27] and magnetic fields[27]. Recently, polaron states have also attracted significant interest as sensitive probes of strongly correlated electronic phases, such as Wigner crystals[28-30], quantum Hall states[20], correlated Mott insulating states emerging in moiré superlattices[31], as well as spin sensors of correlated topological phases[32].

In this work, we scrutinize the optical properties of many-body complexes at the interface between monolayer $MoSe_2$ and epitaxially grown GaInP crystals, both among the most technologically relevant representatives of TMDCs and III/V semiconductors, respectively[4,33]. The optical bandgap of GaInP (lattice-matched to GaAs with the chemical composition $Ga_{0.51}In_{0.49}P$) is sufficiently large to make it compatible with the integration of various TMDC monolayers and bilayers, in stark contrast to, e.g., GaAs or Si. Earlier reports have furthermore hinted that the dielectric properties of GaInP yield high-brightness emission of charged carrier complexes from TMDCs in their proximity[7,15,34,35]. Here, we combine scanning tunneling spectroscopy, photoluminescence (PL), and optical absorption measurements to detail the energy landscape at the GaInP/$MoSe_2$ hybrid interface. We observe unambiguous spectral fingerprints of coherent

Fermi polaron quasiparticles, including significant optical absorption of the charged carrier complex, and the absence of the carrier recoil effect at elevated temperatures.

**Results and Discussion**

The sample structure is shown in Fig. 1a. The GaInP layer is grown on a distributed Bragg reflector (DBR). The MoSe$_2$ monolayers and hexagonal boron nitride (hBN) layers are exfoliated from crystals, and deposited via a standard dry viscoelastic transfer method. See more details of sample preparations in Methods. First, we analyze the band-configuration of the hybrid GaInP/MoSe$_2$ interface that is essential to understand the optical signatures and fingerprints. We therefore conduct low-temperature (100 K) scanning tunneling microscope (STM) experiments under ultra-high vacuum conditions, utilizing an exfoliated MoSe$_2$ monolayer on a GaInP surface. The top inset of Fig. 1a displays a topographic image of the MoSe$_2$ monolayer on GaInP with atomic resolution. The MoSe$_2$ monolayer clearly reveals the hexagonal pattern of the topmost Se plane with 3.2 Å periodicity[36].

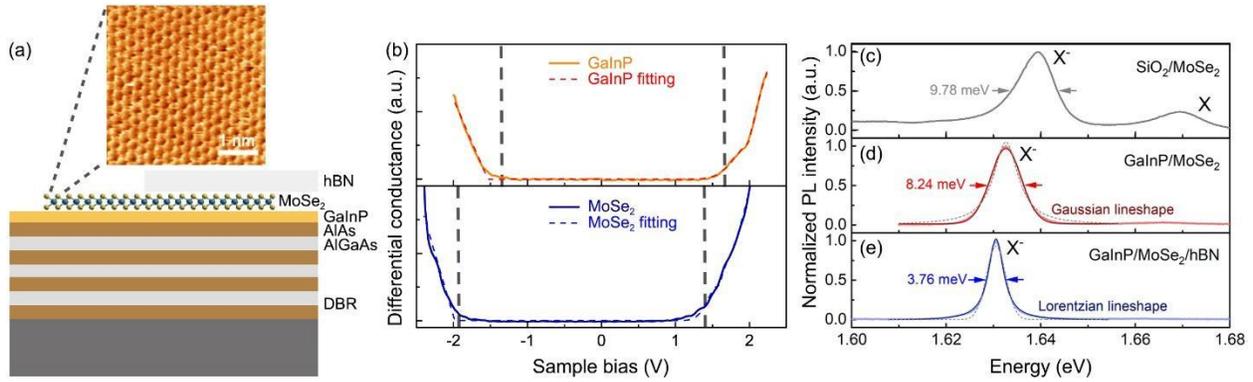

**Figure 1 (a)** Schematic drawing of the investigated GaInP/MoSe$_2$ sample, with an STM tomographic image of MoSe$_2$ monolayer on GaInP surface. **(b)** Differential conductance spectra taken at MoSe$_2$ monolayer and GaInP (Setpoint 2.0 V, 0.1 nA). Normalized photoluminescence spectra recorded at T = 4 K of **(c)** uncovered MoSe$_2$ on a SiO$_2$ substrate, **(d)** uncovered and **(e)** hBN-covered MoSe$_2$ on a GaInP substrate, respectively. **(c)** On SiO$_2$, we observe a pronounced trion peak in photoluminescence (X$^-$) and a weaker exciton peak (X). **(d)** On GaInP, we observe almost no emission from the exciton, while the remaining luminescence peak (X$^-$) has a Gaussian line shape (dark red line). The dashed line represents a Lorentzian fit, underlining the deviation of the measured signal from this lineshape. **(e)** Upon covering the flake with a thin layer of hBN, the emission shifts to slightly lower energies, and the linewidth drops considerably. The spectrum is in almost perfect agreement with a Lorentzian fit (dark blue line). In contrast to **(d)**, the Gaussian fit (dashed line) does not reproduce the spectrum.

Differential conductance (dI/dV) spectra are acquired to probe the band positions of the GaInP substrate and the MoSe$_2$ monolayer with respect to the Fermi level. Example spectra for each surface are depicted in Fig. 1b; they are fitted with linear slopes to determine the band edges and a Gaussian function to account for possible gap states. A statistical average for the band onsets has been derived from several dozens of spectra, and the respective mean values of the valence band (VB) and conduction band (CB) onset are indicated as dashed lines in Fig. 1b. The GaInP shows a nearly symmetric bandgap with VB and CB onsets at -1.3 ± 0.3 V and +1.7 ± 0.3 V, respectively, as well as a surface state at ~1.3 V[37]. We note that the measured band gap is larger than literature values for the $Ga_{0.5}In_{0.5}P$ composition[38], most likely due to band bending effects in the tip electric field. Also, the intrinsic n-type conductance behavior seems to be washed out, probably due to a naturally evolving surface passivation layer. As a comparison, the $MoSe_2$ monolayer has a clearly downshifted bandgap with VB and CB onsets at -1.9 ± 0.2 V and +1.4 ± 0.3 V, respectively. Thus, the detected band alignment corresponds to a type II hetero-interface between GaInP and the $MoSe_2$ monolayer, where the CB confines electrons in $MoSe_2$, leaving holes effectively unconfined. Additionally, we make use of density functional theory (DFT) – a powerful tool to study electronic and optical properties of TMDC monolayers[39, 40] – to confirm a type II band alignment of the GaInP/$MoSe_2$ heterostructure (Fig. S1), in agreement with the result of differential conductance spectra. Particularly, the lowest unoccupied level being localized in the $MoSe_2$ promotes efficient electron transfer across the interface into the TMDC monolayer.

To probe the optical properties, we record low-temperature (4K) PL spectra at different positions on three samples: (i) uncovered $MoSe_2$ monolayer on a $SiO_2$ substrate, (ii) uncovered and (iii) hBN-covered $MoSe_2$ monolayer on a GaInP substrate. Representative spectra for each group are shown in Fig. 1c-e and display significant variations in their response due to the different dielectric environments of the TMDC monolayer. As shown in Fig. 1c, the PL spectrum of $MoSe_2$ on a $SiO_2$ substrate features two prominent broad peaks that can be readily attributed to resonances connected to the exciton and trion of $MoSe_2$ monolayer in the presence of modest charging. The trion peak takes up around 85% of the total emission, suggesting an accumulation of carriers in the TMDC layer. In addition, the lineshape of the trion luminescence fits neither a Gaussian nor Lorentzian profile. This phenomenon can be understood as a result of the electron recoil effect[41-43], as we discuss further below.

In stark contrast, the exciton resonance vanishes almost completely in PL spectra recorded from the $MoSe_2$ monolayer on a GaInP substrate (Fig. 1d). This behaviour, which has been previously reported in Ref.[15], suggests an efficient charge transfer at the GaInP/$MoSe_2$ interface resulting in heavy doping of the TMDC monolayer. The PL signal clearly exhibits a Gaussian lineshape with a full width at half maximum (FWHM) of 8.3 meV, evidencing a reduction of large-scale inhomogeneities compared to the $SiO_2$/$MoSe_2$ interface. The dashed grey line in Fig. 1d shows an alternative Lorentzian fit for comparison, and its poor match to the spectrum underlines the Gaussian distribution of the measured data. Moreover, in comparison to the $SiO_2$/$MoSe_2$ sample, the GaInP/$MoSe_2$ emission signal undergoes a redshift of ~7.6 meV, being explained by the modified dielectric environment and an energy renormalization arising from increased correlations at heavy doping conditions.

Upon placing a thin layer of hBN on top of the $MoSe_2$ flake and annealing it at 400 °C in an Ar:$H_2$ atmosphere (95:5 ratio), we observe further modifications in the optical properties of our samples. Apart from a modest red shift by 0.5 meV, most importantly, we notice a transition to a Lorentzian lineshape, indicating a reduction of inhomogeneities. In Fig. 1e, we include an alternative Gaussian fit (dashed line) to show an inferior match to the measured data and illustrate the almost perfect Lorentzian lineshape of the spectrum. Intriguingly, the recorded optical linewidth narrows down to a value as low as 3.8 meV, corroborating the strongly suppressed inhomogeneous broadening effects. We note that, although narrow-band photoluminescence upon full hBN encapsulation of TMDC monolayers has been observed in many labs[35, 44, 45], such high-quality emission features with exceptionally small FWHM are only sparsely documented in hybrid heterostructures[9].

To back up our findings statistically, we plot the extracted linewidths of the photoluminescence peak from the tentatively assigned charged complex with 0.5 meV binning in a histogram (Fig. 2a). The progressive reduction of the average linewidth becomes immediately evident, when changing from SiO$_2$ to GaInP substrates and applying a hBN cover. For SiO$_2$/MoSe$_2$ and hBN-covered GaInP/MoSe$_2$, the linewidths spread over ~6 meV, while for uncovered GaInP/MoSe$_2$, they span a range of less than 2 meV, with an average value of 8.36 ± 0.44 meV. The enhanced linewidth fluctuation for the hBN-covered GaInP/MoSe$_2$ sample can be explained by the improper contact between TMDCs and hBN cap in certain areas, due to the formation of bubbles during the transfer and annealing procedures. For these sample areas, the PL linewidth approaches the values of the uncapped sample.

The linewidth variance of GaInP/MoSe$_2$ is smaller than that of SiO$_2$/MoSe$_2$ sample. This is in full agreement with a previous work[15], where the linewidth of single photon emission in GaInP/WSe$_2$ has a smaller fluctuation than that of SiO$_2$/WSe$_2$. The reason behind them could be the same: the epitaxial grown substrate GaInP has better surface quality than SiO$_2$ substrate. Charge puddling is known to occur on SiO$_2$ surfaces, promoted by disorders such as dangling bonds and carrier traps, which causes stronger inhomogeneous broadening and linewidth fluctuation. Besides, the dielectric constant of SiO$_2$ is not as uniform as GaInP, contributing to a larger fluctuation of linewidth.

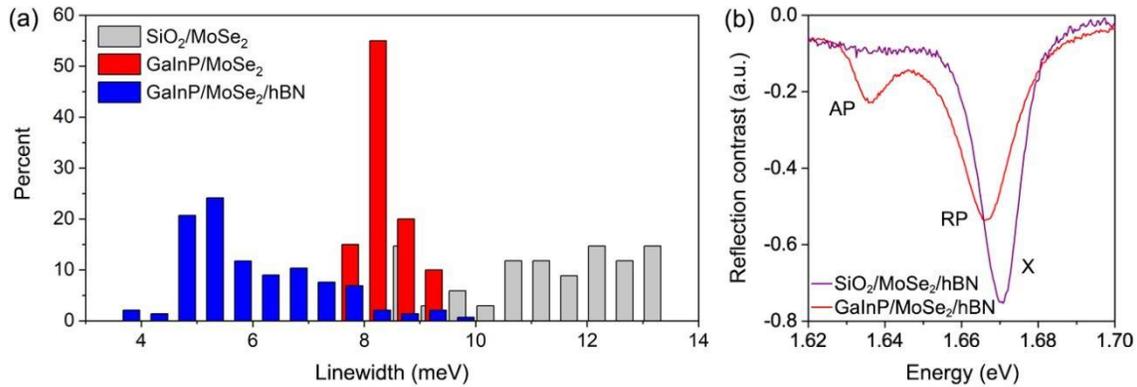

**Figure 2 (a)** Statistical analysis of the linewidth distribution of the MoSe$_2$ PL signal arising from the charged complex measured at different positions on the three samples. The measured linewidths are binned in 0.5 meV. On the SiO$_2$ substrate (gray bars), the linewidth is 11.31 meV in average and varies over a range of 5 meV. On the GaInP substrate (red bars), the linewidths lie within a range of 1.6 meV and the average linewidth is 8.36 meV. After covering the sample with a hBN thin layer (blue bars), the range of linewidths increases again to 5.76 meV, while the average linewidth narrows down to 5.95 meV and the linewidth can possibly reach a value as small as 3.76 meV. **(b)** White light reflection contrast spectra of a hBN-covered MoSe$_2$ flake on GaInP (red) and SiO$_2$ (purple). A strong absorption dip corresponds to the exciton resonance (X) for SiO$_2$/MoSe$_2$/hBN, whereas in GaInP/MoSe$_2$/hBN spectra, a weaker absorption dip with lower energy indicates the emergence of attractive Fermi polaron (AP) complex, and the high energy absorption signal is assigned to the repulsive Fermi polaron (RP) counterpart.

In PL experiments on stochastically doped TMDC monolayers, the emergence of a lower energy peak is usually assigned to the formation of trions. However, the investigation of the oscillator strength via optical absorption allows us to draw more explicit conclusions. For this purpose, we carry out additional reflection measurements at 4 K, as depicted in Fig. 2b. The optical reflectivity of the SiO$_2$/MoSe$_2$/hBN sample displays the standard behavior of ungated samples[14]: While the assigned exciton peak features a very strong absorption signal, the absorption of the trion complex is strongly suppressed. This is in stark contrast to the observed behavior of our GaInP/MoSe$_2$/hBN samples: Here, we consistently observe two optical transitions with significant photo-absorption, whereby the low-energy absorption feature spectrally matches with the PL from the same sample as shown in Fig. 1e. The observation of strong oscillator strength of the charged carrier complex is especially striking, and allows us to assign the carrier complex to an emergent attractive Fermi polaron, with the higher energy absorption signal being its repulsive Fermi polaron counterpart[20, 21].

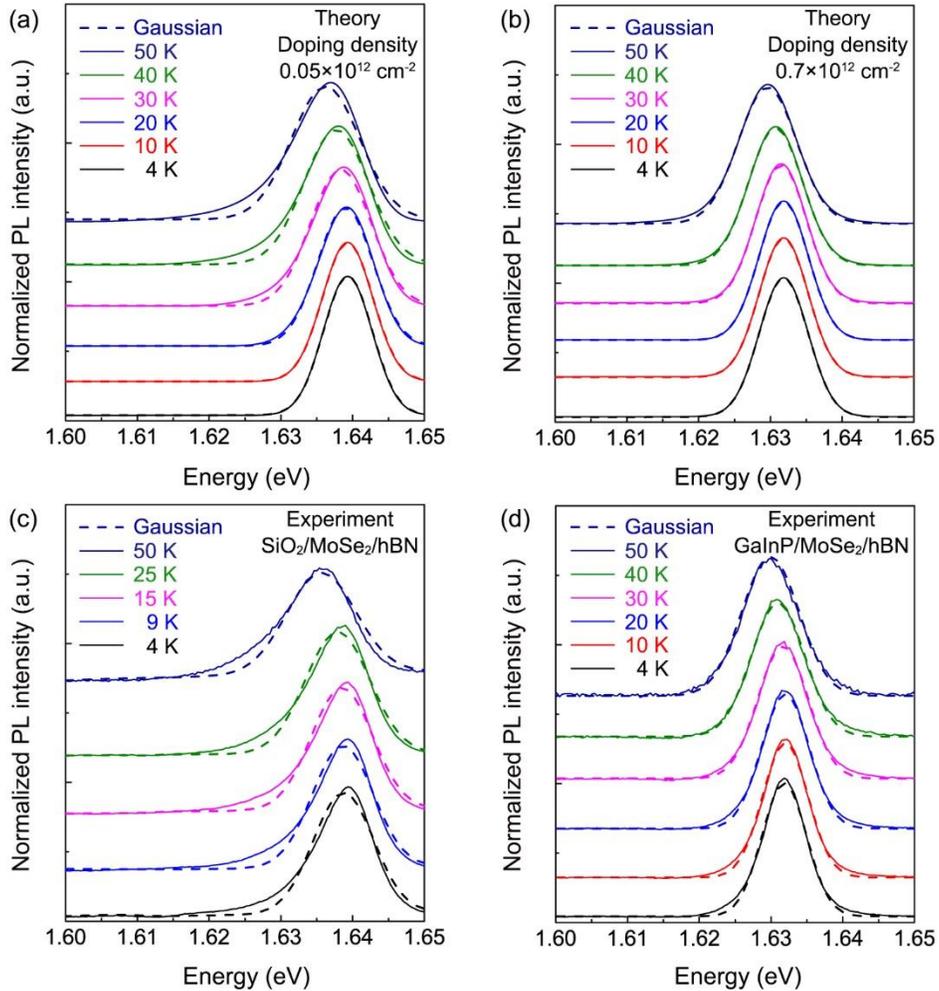

**Figure 3 (a, b)** Temperature evolution of the theoretical PL emission spectra of the charged complex in MoSe$_2$ monolayer evaluated at two different densities: **(a)** $0.05 \times 10^{12}$ cm$^{-2}$ and **(b)** $0.7 \times 10^{12}$ cm$^{-2}$. The doping density in **(b)** has been estimated by comparing the experimental and theoretical temperature dependence of the repulsive and attractive polaron energy splitting for the GaInP/MoSe$_2$/hBN sample (see Fig. S2). The dashed lines are Gaussian fittings to the theoretical profiles. **(c)** Experimental temperature dependent PL of the assigned trion signal in the SiO$_2$/MoSe$_2$/hBN stack, displaying a pronounced asymmetry at finite temperature due to the carrier recoil effect. **(d)** Experimental temperature dependent PL of the assigned attractive polaron signal in the GaInP/MoSe$_2$/hBN stack, retaining its symmetry over the entire temperature range from 4 - 50 K. All dashed curves are Gaussian fittings.

Besides a substantially different absorption behavior, AP complex and bound trion also differ in their temperature-dependent photoluminescence. To elucidate the distinction between attractive Fermi polaron and trion from the PL lineshapes, we calculate the PL within a finite-temperature non-perturbative Green's function approach[46, 47]. Crucially, our theoretical framework captures the crossover[47] between an incoherent trion continuum at low charge densities $E_F \ll k_B T$ and a coherent Fermi-polaron quasiparticle at high densities $E_F \gg k_B T$, where T is the temperature and $E_F$ is the Fermi energy associated with the doping of charges. Here, the attractive Fermi polaron in the quantum-degenerate regime leads to a symmetric emission profile, while at high temperature or low doping the attractive polaron quasiparticle is destroyed, merging with the broad trion-hole continuum. It leads to an asymmetric profile with an exponential tail, involving recoil charge carriers such as electrons at finite momentum. This phenomenon is known as the carrier recoil effect, which has been observed in charge tunable $MoSe_2$ monolayer under modest doping conditions[41].

This is clearly observed in Fig. 3a, b, where we present the lineshape evolution of the charged complex (trion or AP) in the $MoSe_2$ monolayer as a function of temperature at two fixed densities. In Fig. 3a, we fix the density to a small value $0.05 \times 10^{12}$ cm$^{-2}$ expected for the $SiO_2/MoSe_2/hBN$ sample, and we observe an asymmetric exponential tail developing at high temperature. By contrast, in Fig. 3b, the doping density has been fixed to $0.7 \times 10^{12}$ cm$^{-2}$. This corresponds to the carrier density expected in the $GaInP/MoSe_2/hBN$ sample and has been estimated by comparing the temperature dependence of the repulsive and attractive polaron energy splitting (see Supplementary information Fig. S2). Here, we observe that, over the entire temperature range considered, the emission PL profile of the attractive Fermi polaron resonance remains symmetric and can be well fitted with Gaussians (dashed lines), indicating that the attractive polaron persists as a well-defined quasiparticle. We also calculated the reliance of PL lineshape on the doping density at 50 K, and find the suppression of recoil effect with doping (Fig. S3).

Temperature dependent PL spectra assigned to trion emission in the $SiO_2/MoSe_2/hBN$ sample are shown in Fig. 3c. All spectra are normalized with respect to their maximum intensity. As we ramp the temperature from 4 K to 50 K, a distinct low-energy tail becomes visible at the emission peaks for the case of trion photoluminescence. The strong deviation from a Gaussian fit (dashed line) is evident at 50 K. In striking contrast, the AP PL from the $GaInP/MoSe_2/hBN$ sample as exhibited in Fig. 3d, only shows a slight spectral broadening, without any signs of asymmetry as temperature increases. This is evidenced by the excellent match to a Gaussian fit at 50 K, implying the disappearance of the recoil effect. Thus, besides the absorption characteristics (see also Fig. S2), the temperature-dependent PL provides another clear distinction between trions and AP.

**Conclusions**

Our study reveals the formation of attractive Fermi polarons at the $GaInP/MoSe_2$ hybrid interface, which emerges as the dominant feature in photoluminescence, with a characteristic dependence on temperature. Due to the integration of TMDC monolayers with epitaxially grown surfaces, we observe a strong reduction of dielectric disorder, giving rise to enhanced optical properties, such as luminescence with strongly suppressed disorder-related broadening. We demonstrate a considerable charge-doping mechanism at the $GaInP/MoSe_2$ interface, which results in a significant transfer of optical oscillator strength to the attractive Fermi polaron resonance. We show that, in contrast to the low-doping regime, the low-energy carrier-recoil exponential tail is suppressed in temperature-dependent photoluminescence, consistent with the attractive polaron remaining a well-defined quasiparticle at temperatures below the Fermi energy. Furthermore, the presence of both repulsive and attractive polarons in reflectance, along with the temperature dependence of their energy splitting, matches the predicted behavior of polaron quasiparticles. Our findings are important for engineering mixtures of excitons and charge carriers in hybrid structures, e.g., in combined optics-transport experiments.

**Methods**

**Sample and experimental setup**

The MoSe$_2$ monolayers were exfoliated from crystals grown by chemical vapor deposition, and deposited on the substrates via a standard dry viscoelastic transfer method[48]. The SiO$_2$ substrates used in this study consist of a 90 nm-thick SiO$_2$ layer on top of a silicon wafer. The GaInP substrates were made of a 10 nm-thick lattice-matched Ga$_{0.51}$In$_{0.49}$P layer grown on an AlAs/AlGaAs distributed Bragg reflector (DBR) via molecular beam epitaxy (Fig. 1a). The DBR not only enhances the light output, but also allows for straightforward absorption measurements in the back-reflection geometry. Substrates used for tunneling spectroscopy were grown via the same method, but lack the DBR to enhance their electric conductivity. Additionally, we produced a sample where the MoSe$_2$ flake was covered with a thin layer (~10 nm) of hBN and subsequently annealed at 400 °C in an Ar:H$_2$ atmosphere (95:5 ratio) for 2 hours. The STM experiments were performed in an ultrahigh vacuum chamber at 100 K, using chemically etched Au tips. Differential conductance spectroscopy was performed with a lock-in amplifier at 1356 Hz and 14 mV modulation frequency. The samples were degassed at 500 K for 1h prior to measurements.

**Computational Details**

To compare with and interpret the experimental PL profiles shown in Fig. 3, we employ a finite-temperature Fermi polaron theory[46, 47] that goes beyond trion-based theories[41-43]. Our theory assumes uncorrelated thermalized excitons and a thermal distribution of charge carriers. Within these approximations, the photoluminescence, $PL(\omega)$, at frequency $\omega$ (measured from the exciton) is proportional to the exciton spectral function via the Kennard-Stepanov relation[49-51], $PL(\omega) = e^{-\beta\omega} A(\omega)$. Here, the exciton spectral function $A(\omega) = -(1/\pi) Im\, G(\omega + i0^+)$ depends on the exciton Green's function $G(\omega) = [\omega - \Sigma(\omega)]^{-1}$ (we work in units where the reduced Planck constant is 1). The details of the many-body dressing of excitons by the charge carriers are contained within the self energy $\Sigma(\omega)$, and, for PL close to the exciton and trion, this is in turn dominated by processes involving at most a single carrier together with the exciton. Within this approximation we have $\Sigma(\omega) = \sum_{\mathbf{k}} f_{\mathbf{k}}\, T(k, \omega + \epsilon_{\mathbf{k}})$, with $k$ the carrier momentum, $\epsilon_k = k^2/(2m)$ its dispersion, $\epsilon_k^X = k^2/(2m_X)$ the exciton dispersion, and $f_k$ the Fermi-Dirac distribution of charges at the given temperature. Importantly, the distribution function in the self energy is responsible for the characteristic recoil effect that occurs at either low doping or high temperature, in agreement with trion-based theory[41-43]. The low-energy scattering $T$ matrix of excitons and electrons close to the trion resonance is given by $T^{-1}(\mathbf{q}, \omega) = -\sum_{\mathbf{k}} \frac{1}{\epsilon_T + \epsilon_{\mathbf{k}} + \epsilon_{\mathbf{k}}^X} - \sum_{\mathbf{k}} \frac{1 - f_{\mathbf{k}}}{\omega - \epsilon_{\mathbf{k}} - \epsilon_{\mathbf{k-q}}^X + i0^+}$ [46, 47]. The theoretical PL spectra are convolved with a Lorentzian of FWHM linewidth of $8\ meV$, using $\epsilon_T = 22.5\ meV$, and the MoSe$_2$ values of the exciton and carrier (hole) effective masses: $m_X = 1.15 m_0$ and $m = 0.59 m_0$ with $m_0$ the free electron mass. The exciton energy at zero temperature and vanishing doping is treated as a fitting parameter.

**Acknowledgement**

The authors gratefully acknowledge funding by Ministry of Research and Culture of Lower Saxony (MWK) within the project DyNano and the Wissenschaftsraum ELiKo. Support by the German Research Foundation (DFG) within the program INST 184/234-1 FUGG is acknowledged. C.C. and A.M.V. acknowledge additional support from MWK (Professorinnenprogramm für Niedersachsen) as well as from the German Ministry of Education and Research (Professorinnenprogramm III). The computational resources were provided by the computing cluster ROSA at the Carl von Ossietzky University of Oldenburg financed by MWK and by DFG (INST 184/225-1 FUGG). S.H. is grateful for financial support by the DFG (HO 5194/16-1). S.T acknowledges support from Applied Materials Inc. (materials synthesis), NSF CBET 2330110 (materials environmental testing), DOE-SC0020653 (TMDCs excitonic testing studies), and Lawrence Semiconductors (precursor purification). K.W. and T.T. acknowledge support from the JSPS KAKENHI (Grant Numbers 21H05233 and 23H02052) and World Premier International Research Center Initiative (WPI), MEXT, Japan. BCM, JL and MMP acknowledge support from the Australian Research Council through Discovery Projects DP240100569 (JL) and 250103746 (BCM, JL, MMP), and Future Fellowship FT200100619 (MMP). FMM acknowledges financial support from the Spanish Ministry of Science, Innovation and Universities through the "Maria de Maetzu" Programme for Units of Excellence in R&D (CEX2023-001316-M) and from the Ministry of Science, Innovation and Universities MCIN/AEI/10.13039/501100011033, FEDER UE, project No. PID2023-150420NB-C31 (Q).